\documentclass[draft,jgrga]{agutex}

 \usepackage{lineno}

  \usepackage[dvips]{graphicx}
  \usepackage{amssymb}
  \setkeys{Gin}{draft=false}
\authorrunninghead{E. CAMPOREALE, A. CAR\`{E} AND J. BOROVSKY}

\titlerunninghead{Solar Wind Classification}

\authoraddr{Corresponding author: E. Camporeale (e.camporeale@cwi.nl)}

\begin{document}

%
%

\title{Classification of Solar Wind with Machine Learning}
%

%

%
%



 \authors{Enrico Camporeale\altaffilmark{1}, Algo Car\`{e}\altaffilmark{1}, Joseph E. Borovsky\altaffilmark{2}}

\altaffiltext{1}{Center for Mathematics and Computer Science (CWI), Amsterdam, The Netherlands}
\altaffiltext{2}{Center for Space Plasma Physics, Space Science Institute, Boulder, Colorado, USA}

%
%

\begin{abstract}
We present a four-category classification algorithm for the solar wind, based on Gaussian Process. The four categories are the ones previously adopted in \citet{xu2015}: ejecta, coronal hole origin plasma, streamer belt origin plasma, and sector reversal origin plasma. The algorithm is trained and tested on a labeled portion of the OMNI dataset. It uses seven inputs: the solar wind speed $V_{sw}$, the temperature standard deviation $\sigma_T$, the sunspot number $R$, the $f_{10.7}$ index, the Alfven speed $v_A$, the proton specific entropy $S_p$ and the proton temperature $T_p$ compared to a velocity-dependent expected temperature. 
The output of the Gaussian Process classifier is a four element vector containing the probabilities that an event (one reading from the hourly-averaged OMNI database) belongs to each category. The probabilistic nature of the prediction allows for a more informative and flexible interpretation of the results, for instance being able to classify events as 'undecided'.
The new method has a median accuracy larger than $90\%$ for all categories, even using a small set of data for training. The Receiver Operating Characteristic curve and the reliability diagram also demonstrate the excellent quality of this new method.
Finally, we use the algorithm to classify a large portion of the OMNI dataset, and we present for the first time transition probabilities between different solar wind categories. Such probabilities represent the 'climatological' statistics that determine the solar wind baseline.
\end{abstract}

%
%

\begin{article}

\section{Introduction}\label{sec:intro}
There is a growing consensus that the classification of the solar wind merely based on the particle velocity is too simplistic \citep{zurbuchen02, salem03, zurbuchen07, borovsky12, stakhiv15}.
Recent classifications distinguish the solar wind based on its solar origin. Recently, \citet{xu2015} have introduced a four-plasma categorization scheme, that distinguishes between ejecta, coronal hole origin, sector reversal origin, and streamer belts origin. A correct classification of the solar wind is important, on one hand to be able to understand and interpret solar wind observations and to analyze data consistently (for instance a statistical study of heterogeneous solar wind types might lead to erroneous conclusions) \citep{wimmer06, borovsky12b, borovsky16, kilpua16, neugebauer16}, and on the other hand to be able to develop space weather models that take into account the different geoeffectiveness of different types of solar wind \citep{crooker77, chen97,gopalswamy15, wing16, riley17}. 
Clearly, along with a correct classification, we need to develop a robust and accurate scheme to {automatically assign} categories. {In this paper we deal with the problem of now-casting the correct classification using solar wind properties taken from the OMNI database.}
\citet{xu2015} have introduced a simple scheme based on only three variables. That scheme produces very good results, particularly for the ejecta and the coronal hole origin category, but less so for streamer belt and sector reversal origin plasmas.
A close inspection of Figure 6 in \citet{xu2015} shows that these two categories have many overlapping data points in the chosen three-dimensional space. Hence, one can argue that an improvement in {classification skill} would almost certainly require an increase in the number of input variables. 
The goal of this paper is to introduce a new scheme that has excellent accuracy for all categories. \\
We employ a Machine Learning technique called Gaussian Process (GP), for a multi-category classification task. GP is a non-parametric statistical model, which has the advantage of producing probabilistic {results, employing a Bayesian approach}. This is a key feature that makes GP models different from most Machine Learning schemes. In particular, other classification schemes have the goal of finding the category boundaries in a multidimensional input space, unambiguously partitioning the input space into categories. The probabilistic approach is more flexible and informative because it attaches a confidence level to predictions, hence being able to classify events as 'undecided', if needed.

\section{Data}\label{sec:data}
We use the same data that was utilized in \citet{xu2015}.
Xu and Borovsky created four collections of data that they considered to be known examples of the four types of solar-wind plasma: (1) magnetic clouds as clear examples of ejecta, (2) unperturbed high-speed streams as clear examples of coronal-hole-origin plasma, (3) strahl-confusion zones as clear examples of sector-reversal-region plasma, and (4) pseudostreamers as clear examples of streamer-belt-origin plasma, which we denote hereinafter as categories 1 to 4, respectively. 
{The magnetic clouds as clear examples of ejecta and the unperturbed high-speed streams as clear examples of coronal-hole-origin plasma need little explanation, but the strahl-confusion zones and the pseudostreamers do need explanation. Around magnetic sector reversals (i.e. the heliospheric current sheet) there can be a distinct plasma that has been denoted as the heliospheric plasma sheet \citep{winterhalter94}, however \citet{xu2015} found that some sector-reversal regions have such plasma-sheet plasma and some do not. They did consistently find dropouts of the solar-wind heat flux (strahl) in the vicinity of the heliospheric plasma sheet and determined that the heat-flux dropouts (or “strahl confusion zones”) were a better marker of the plasma in the vicinity of the sector reversals. Sometimes a solar-wind spacecraft traverses from one coronal-hole region to another coronal-hole region of the same magnetic polarity, i.e. without crossing a magnetic-sector reversal; this is known as crossing a pseudostreamer. The slow solar wind between those coronal holes is of streamer-belt origin without contamination of sector-reversal-region plasma. Hence the slow wind during pseudostreamer crossings is a clear example of streamer-belt-origin plasma.}\\
After removing some gaps in the data, our dataset is composed of 1925 {(21.5\%)} 1-hour events categorized as ejecta, 3049 {(34\%)} coronal hole origin, 2273 {(25.5\%)} sector reversal, 1704 {(19\%)} streamer belts, for a total of 8951 data points. The dataset comprises events from 1995 to 2008. The attributes are the same contained in the low resolution OMNI web dataset (https://omniweb.gsfc.nasa.gov/), and one event is intended as one reading from the hourly averaged OMNI database.
In \citet{xu2015} a classification algorithm was proposed that makes use of only three variables, namely the Alfven speed $v_A$, the proton specific entropy $S_p=T_p/n_p^{2/3}$ (with $T_p$ and $n_p$ the proton temperature and density, respectively), and the temperature ratio $T_{exp}/T_p$, where $T_p$ is the measured proton temperature, and $T_{exp}$ is the expected temperature of protons, calculated as:
\begin{linenomath*}
\begin{equation}
 T_{exp} = (V_{sw}/258)^{3.113},
\end{equation}
\end{linenomath*}
with $V_{sw}$ the solar wind speed in km/s.
Despite its simplicity, the three-parameters four-categories scheme proposed in \citet{xu2015} yields a remarkable accuracy. It categorizes correctly 87.5\% and 96.9\% of the ``known'' events belonging to categories 1 and 2, respectively. It performs less well for categories 3 and 4, where it scores 69.9\% and 72.0\%. These are the scores that we aim to improve in this paper.

\section{Method}\label{sec:method}
We use a Gaussian Process (GP) classification method. For a detailed technical description, we refer the reader to standard textbooks, e.g. \citep{rasmussen2006, bishop2007}. This is a statistical model that is generally defined as a probability distribution over functions $y(\mathbf{x})$ (with $\mathbf{x}$ a multidimensional variable) such that the set of values ${y_1,y_2,\dots,y_N}$ evaluated at any set of points $\mathbf{x}_1, \mathbf{x}_2,\dots,\mathbf{x}_N$ jointly have a Gaussian distribution. As such, the joint distribution is completely determined by its second-order statistics, namely mean and covariance. Here, we can assume without loss of generality that the data is rescaled to have zero mean, and thus the Gaussian Process is determined by the specification of the covariance of $y(\mathbf{x})$ defined as
\begin{linenomath*}
\begin{equation}
 \mathbb{E}[y(\mathbf{x}_1)y(\mathbf{x}_2)] = k(\mathbf{x}_1,\mathbf{x}_2),
\end{equation}
\end{linenomath*}
where the function $k$ is known as kernel or covariance function. This defines the similarity between different data points, and it encodes all the assumptions made in the Gaussian process model.
A Gaussian Process allows to perform a probabilistic classification, where a prediction is interpreted as the probability that an event belongs to a given class. 
In a Gaussian Process used for binary classification, it is customary to squash the output of a GP regression model from the real axis to the interval $[0,1]$, using a sigmoid. Here, we will employ the logistic function $\pi(f) = \frac{1}{1+\exp(-f)}$. In this way, $\pi$ has a natural probabilistic interpretation. In practice, one assumes a Gaussian distribution of so-called \emph{latent} functions $f(\mathbf{x})$, that are mapped through the sigmoid $\pi$. In turn, this produces a non-Gaussian Process
over the functions $\pi$ (interpreted as probabilities). The values taken by $f$ however are not observable and not of interest. In order to obtain a predictive distribution at a new point $\mathbf{x}^*$, the probability $\pi$ must be marginalized over the values of the latent function $f$. Such integral is analytically intractable, and must be approximated \citep{williams98}.
In this work we use an approach called Laplace approximation, where essentially the non-Gaussian distribution of the latent functions $f$, conditioned on the observed data, is approximated by a Gaussian \citep{bishop2007}.
All the results of this paper have been obtained with the MATLAB software GPML, available at http://www.gaussianprocess.org/gpml/code \citep{rasmussen10}.

\subsection{Choice of the kernel and attributes}\label{attributes}
We have used seven attributes, either obtained directly from the OMNI database, or as derived quantities. Specifically, we use: the solar wind speed $V_{sw}$, the temperature standard deviation $\sigma_T$ {calculated on 1-hour intervals}, the sunspot number $R$, the $f_{10.7}$ index, the Alfven speed $v_A$, the proton specific entropy $S_p$ and the temperature ratio $T_{exp}/T_p$ (see Table 1). The last three are the same attributes used in \citet{xu2015}. 

\begin{table}[!ht]
 \caption{List of attributes}\label{table:attributes}
 \centering
 \begin{tabular}{cc}
  \hline
  Attribute & Symbol\\
  \hline
  Solar wind speed & $V_{sw}$\\ 
  Proton temperature standard deviation & $\sigma_{T}$\\
  Sunspot number & $R$\\
  Solar radio flux (10.7 cm) & $f_{10.7}$\\
  Alfven speed & $v_{A}$\\
  Proton specific entropy & $S_p$\\
  Temperature ratio & $T_{exp}/T_{p}$\\
  \hline
  
 \end{tabular}

\end{table}

\begin{figure}[ht!]
 \centering
 \includegraphics[width=30pc]{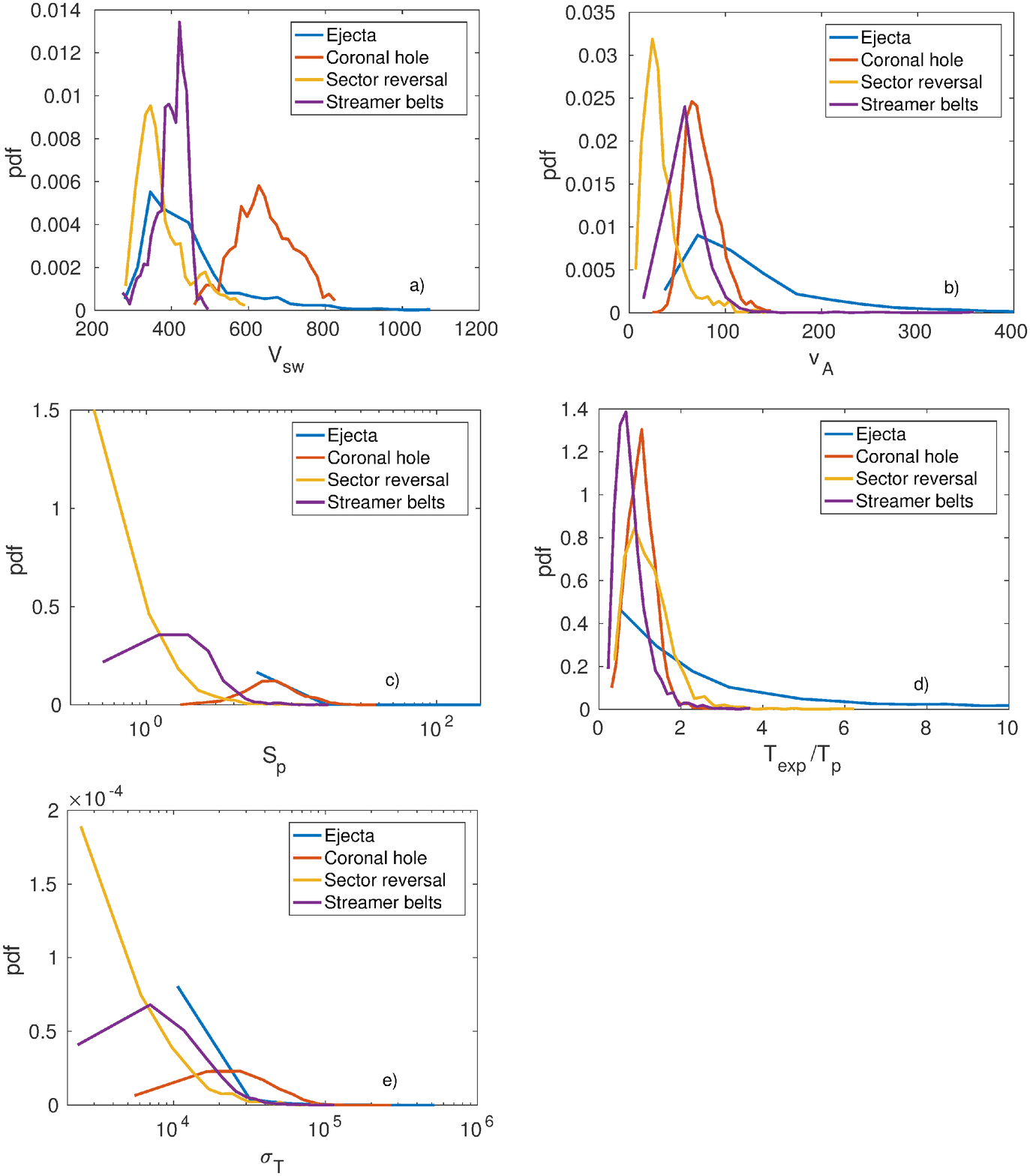}
 \caption{Probability distribution function (pdf) of the attributes $V_{sw}$, $v_A$, $S_p$, $T_{exp}/T_p$, $\sigma_T$ respectively in panel a) to e), shown for the four different categories, calculated over the whole training set. The pdfs are normalized such that the area under the curve is equal to one.}\label{fig:pdf}
\end{figure}

The attributes have been chosen mostly by trial and error, somewhat guided by physical intuition.
We have verified that using a subset of these seven attributes, the performance is either comparable or worse.
Interestingly, although $R$ and $f_{10.7}$ are known to be strongly correlated, we have verified that using both attributes the performance significantly increases, with respect to the case when one of the two is left out.
{Figure \ref{fig:pdf} shows the probability distribution of five attributes, calculated over the whole database, and colour coded for different categories. $R$ and $f_{10.7}$ are not shown because they do not present any meaningful distribution, for any category.}\\

A crucial ingredient in the GP methodology is an appropriate choice of the covariance function (kernel). The strategy followed here is to use one that belongs to a fairly general family, that is then optimized (see later about ARD) in order to be representative of the data.
The kernel is stationary, meaning that $k(\mathbf{x},\mathbf{y})$ is a function of $r=||\mathbf{x}-\mathbf{y}||$ only. It is chosen as a combination of an isotropic Gaussian and a piecewise polynomial kernel with compact support:
\begin{linenomath*}
\begin{equation}
 k(\mathbf{x},\mathbf{y}) = \sigma_0^2\exp\left(-\frac{r^2}{2l_0^2}\right) + \sigma_1^2(1-r_\Sigma)^6(6r_\Sigma+1) \Theta(1-r_\Sigma),
\end{equation}
\end{linenomath*}
where $\Theta$ is the Heaviside step function, the weighted distance $r_\Sigma$ is defined as
\begin{linenomath*}
\begin{equation}
 r_\Sigma = \sqrt{(\mathbf{x}-\mathbf{y})^T\Sigma^{-1}(\mathbf{x}-\mathbf{y})},
\end{equation}
\end{linenomath*}
and $\Sigma$ is the diagonal matrix $\Sigma_{ii}=l_i^2$, with $i=1,\ldots,7$.
Note that the Gaussian part of the kernel can in principle correlate inputs whose distance is larger than $r=1$, while the polynomial is restricted by the Heaviside function to $r_\Sigma<1$. Also, the former assigns an equal length-scale $l_0$ to each input, while the latter implements Automatic Relevance Determination (ARD) \citep{mackay96, neal12}, where different inputs are assigned different $l_i$. Those length-scales define the characteristic distance that determine un-correlation between data. 
The kernel so defined contains 10 hyperparameters: two scaling factors $\sigma$, and eight characteristic length-scales $l$. These are determined by minimizing the negative log likelihood of the training set \citep{rasmussen2006}. In this way, the kernel is tuned to be representative of the training data. 

\section{Results}
A binary classification algorithm can be generalized to multiple categories in a straightforward way. First, we compute the probability of an event to belong to a single category via binary classification, that is 'one category versus all the others'. 
The obtained probabilities can then be re-normalized in order to obtain the correct probability that an event belongs to any of the four categories. Finally, in order to assign an event to a single category, we choose the one with highest probability (however see later for a discussion on the possible choice of thresholds).\\
We have considered cases where 10\%, 15\%, 20\%, and 25\% of the original data (8951 1-hour events) are used for training, and the rest for testing.
It is important to perform cross-validation, that is to make sure that our results do not depend on a particular fortuitous choice of the training set. {In order to do so, we perform 100 independent runs of classifications, each time randomly partitioning the original ``known-event'' set into training and test sets.}
Figure \ref{fig:histogram} shows the accuracy of the classification (percentage of ``known'' events classified correctly) calculated over the distribution of 100 runs. 
The different colors represent the different ratios of training set: from 10\% to 25\% going from lighter to darker orange. The boxes denote the first and third quartile. Horizontal lines and circles show the median and mean values, respectively. Finally, the whiskers denote the $2^{nd}$ and $98^{th}$ percentile of the distribution. 
One can appreciate that the classifier yields extremely good results for all categories: even with only $10\%$ of the data used for training, the median and mean values are above 0.9 for all categories. This result considerably improves the classification scheme proposed in \cite{xu2015}. Similarly to the results in \cite{xu2015}, the coronal hole origin plasma is the easiest to classify, with almost perfect classification. Table \ref{table:accuracy} reports the numerical values of the medians shown in Figure \ref{fig:histogram}. 

\begin{table}[!ht]
\caption{Median percentage of number of events correctly categorized over 100 tests, for different ratios of training to test data size.}\label{table:accuracy}
\centering
\begin{tabular}{c c c c c}
\hline
\multicolumn{5}{r}{Ratio of training to test data}\\
 Category  & 10\% & 15\% & 20\% & 25\% \\
\hline
  Ejecta  &  90.8 & 93.1 & 94.9 & 96.1\\
  Coronal hole origin  & 99.2 & 99.5 & 99.7 & 99.7\\
  Sector reversal origin   & 92.2 & 95.1 & 96.9 & 97.5 \\
  Streamer belt origin  & 93.2  & 96.8 & 97.8  & 98.7 \\
\hline
\end{tabular}
\end{table}

{Table \ref{table:confusion_1} shows the confusion matrix for the case of 20\% training set. The number reported here are the probabilities of predicting a given category, conditioned on the observed categories. This means that the probabilities sum to 100\% along each column. For completeness, we report the numbers of events for each observed categories used to define these probabilities: ejecta (147,700), coronal hole origin (260,100), sector reversal origin (125,600), streamer belt origin (182,500). The percentage of Table \ref{table:confusion_1} were calculated by using the category with maximum probability as the predicted one. However, the probabilistic nature of the Gaussian Process approach allows us to give more informative statistics, depending on the values of the outcome probabilities. This will the subject of the next Section. 

\begin{table}[!ht]
\caption{Confusion matrix for the case of 20\% training set. Probabilities are conditioned on the observed category.}\label{table:confusion_1}
\centering
\begin{tabular}{c c c c c}
\hline
\multicolumn{4}{r}{Observed category}\\
 Prediction  & Ejecta & Coronal hole & Sector reversal & Streamer belt\\
\hline
  Ejecta  &  94.8 & 0.2 & 1.6 & 1.4\\
  Coronal hole origin  & 0.8 & 99.6 & 0.5 & 0.2\\
  Sector reversal origin   &1.8 & 0.1 & 96.6 & 0.8 \\
  Streamer belt origin  & 2.5 & 0.1 & 1.3 & 97.7\\
\hline
\end{tabular}
\end{table}

Table \ref{table:confusion_2} reports the confusion matrix, when only results with largest probability greater than 0.5 are taken in consideration (again only for the case of 20\% training set). With this particular choice, one would disregard $17\%, 3\%, 6\%, 6\%$ of the results (respectively for observed categories 1 to 4), but the frequency of false classification becomes very small.} 

\begin{table}[!ht]
\caption{Confusion matrix for the case of 20\% training set, when only probabilities larger than 50\% are considered. Probabilities are conditioned on the observed category.}\label{table:confusion_2}
\centering
\begin{tabular}{c c c c c}
\hline
\multicolumn{4}{r}{Observed category}\\
 Prediction  & Ejecta & Coronal hole & Sector reversal & Streamer belt\\
\hline
  Ejecta  &  97.9 & 0 & 0.8 &0.6\\
  Coronal hole origin  & 0.2 & 100 & 0.1 & 0.1\\
  Sector reversal origin   &1.0 & 0.0 & 98.5 & 0.3\\
  Streamer belt origin  & 1.0 & 0.0 & 0.5 & 99.0\\
\hline
\end{tabular}
\end{table}

\subsection{Receiver Operating Characteristic (ROC) curve}
\begin{figure}[ht!]
 \centering
 \includegraphics[width=30pc]{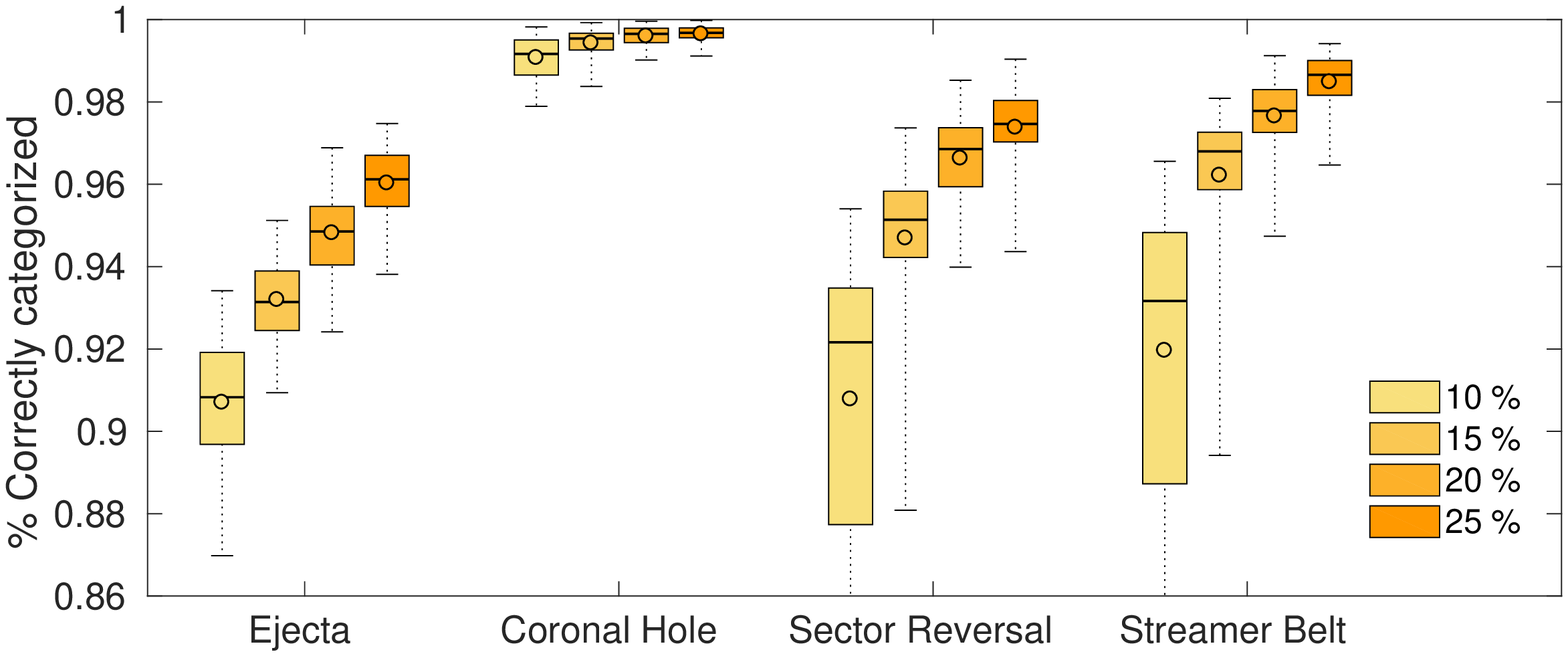}
\caption{Accuracy of the GP classification model. The four colors represent different size of the training set, increasing from light orange (10\% of the dataset used for training) to dark orange (25\%).
The statistics is computed on the percentage of the number of events classified correctly, calculated across 100 runs (the training set being chosen randomly for each run). The boxes denote the $1^\textrm{st}$ and $3^\textrm{rd}$ quartile of the distribution. Horizontal lines and circles show the median and mean values, respectively. The lower and upper whiskers denote the $2^\textrm{nd}$ and $98^\textrm{th}$ percentile.}
\label{fig:histogram}
\end{figure}

Because our classifier produces a probabilistic classification, one is faced with the problem of how to interpret such probabilities. Of course, one possibility is to disregard the probabilistic nature and simply consider the category with higher probability, as done in Figure \ref{fig:histogram}. On the other hand, it is interesting to evaluate how considering as positives only the events that yield probabilities above a certain threshold can change the classification accuracy, in terms of true and false positives and negatives. 
A concise representation of this metric for binary classification is given by the Receiver Operating Characteristic (ROC) curve. This is a curve of the False Positive Ratio $FPR$, defined as the ratio of false positives divided by the total number of negatives (also called 'one minus specificity') versus the True Positive Ratio $TPR$, defined as the ratio of true positives divided by the total number of positives (also called sensitivity), for different values of thresholds.  By definition,  considering a threshold equal to $0\%$ yields $FPR=TPR=1$ and considering a threshold equal to 100\% yields $FPR=TPR=0$. Hence, increasing the threshold designs a well defined curve in the $(FPR,TPR)$ space. A perfect classification would yield $FPR=0$ and $TPR=1$, hence the value of the threshold that produces the point closest to $(FPR=0,TPR=1)$ is optimal. On the other hand a classifier that assigns a pre-fixed probability to all events would produce $FPR=TPR$, thus the distance from this line denotes the quality of a scheme.

Tables \ref{table:roc_ejecta}-\ref{table:roc_st} list the $FPR$ and $TPR$ for each category and thresholds at 10\% intervals (here we have used the scheme with 20\% used for training). The values denoted in bold indicate the optimal threshold, calculated as the one that minimizes the Euclidean distance from $(FPR=0, TPR=1)$.
Here, each table shows the ROC computed in a binary classification sense, i.e. 'one category vs all others'.
Although the ROC is usually shown graphically as a curve in the $(FPR,TPR)$ space, one can appreciate that this model produces such excellent results, with the optimal value extremely close to the point $(FPR=0, TPR=1)$, that plotting the curves would not be very informative.
\begin{table}[!ht]
\caption{False and true positive ratios for category ejecta}\label{table:roc_ejecta}
\centering
\begin{tabular}{c c c}
\hline
Threshold & FPR & TPR\\
\hline
10\% &    0.4541 &   0.9975\\
20\% &    0.0922 &   0.9901\\
\bf{ 30}\% &    \bf{0.0255} &  \bf{ 0.9648}\\
40\% &   0.0099 &   0.9048\\
50\% &   0.0037 &   0.8036\\
60\% &   0.0014 &   0.6226\\
70\% &   0.0006 &   0.3669\\
80\% &   0.0003 &   0.1405\\
90\% &   0.0001 &   0.0240\\
\end{tabular}
\tablenotetext{}{The optimal threshold is in bold.}
\end{table}

\begin{table}[!ht]
\caption{False and true positive ratios for category coronal hole origin}\label{table:roc_ch}
\centering
\begin{tabular}{c c c}
\hline
Threshold & FPR & TPR\\
\hline
10\% &  0.1857 &   0.9999\\
20\% &    0.0315 &   0.9992\\
{\bf 30\%} & {\bf   0.0089} &  {\bf  0.9971}\\
40\% &    0.0036 &   0.9902\\
50\% &   0.0011  &  0.9671\\
60\% &    0.0001 &   0.8777\\
70\% &    0.0000 &   0.6652\\
80\% &        0  &  0.3363\\
90\% &         0  &  0.0666\\
\end{tabular}
\tablenotetext{}{The optimal threshold is in bold.}
\end{table}

\begin{table}
\caption{False and true positive ratios for category sector reversal origin}\label{table:roc_ps}
\centering
\begin{tabular}{c c c}
\hline
Threshold & FPR & TPR\\
\hline
10\% & 0.5793  &  0.9981\\
20\% &    0.1166 &   0.9946\\
\bf{30\%} &    \bf{0.0222} &    \bf{0.9850}\\
40\% &    0.0079 &   0.9653\\
50\% &    0.0029 &   0.9240\\
60\% &    0.0012 &   0.8334\\
70\% &    0.0005 &   0.6078\\
80\% &    0.0001 &   0.2482\\
90\% &    0.0000  &  0.0327\\
\end{tabular}
\tablenotetext{}{The optimal threshold is in bold.}
\end{table}

\begin{table}
\caption{False and true positive ratios for category streamer belt origin}\label{table:roc_st}
\centering
\begin{tabular}{c c c}
\hline
Threshold & FPR & TPR\\
\hline
10\% & 0.4967 &   0.9999\\
20\% &    0.1474 &   0.9985\\
30\% &    0.0317 &   0.9935\\
\bf{40\%} &    \bf{0.0101} &    \bf{0.9771}\\
50\% &    0.0034 &   0.9323\\
60\% &    0.0014 &   0.8106\\
70\% &    0.0005 &   0.5253\\
80\% &    0.0002 &   0.2141\\
90\% &    0.0001 &   0.0613\\
\end{tabular}
\tablenotetext{}{The optimal threshold is in bold.}
\end{table}

{Finally, we report in  Table \ref{table:undecided}  the number of undecided events, for different thresholds (considered as the maximum probability over the four categories larger than a certain number), and their classification skill over all categories. Table \ref{table:undecided} has been computed on the 20\% training set case.}

\begin{table}[!ht]
\caption{Percentage of undecided events, for a given threshold (maximum probability along the four categories), and corresponding classification skill (percentage of events correctly classified)}\label{table:undecided}
\centering
\begin{tabular}{c c c}
\hline
 Threshold (\%) & Undecided (\%) & Correctly classified (\%)\\
\hline
 0.3 & 0.2 & 97.6 \\
 0.4 & 1.8 & 98.2 \\
 0.5 & 7.5 & 99.1 \\
 0.6 & 19.7 & 95.6 \\
 0.7 & 44.1 & 95.8 \\
 0.8 & 75.1 & 99.8 \\
 0.9 & 95.0 & 100 \\
 \hline
\end{tabular}
\end{table}

\subsection{Reliability}
\begin{figure}[ht!]
 \centering
 \includegraphics[width=30pc]{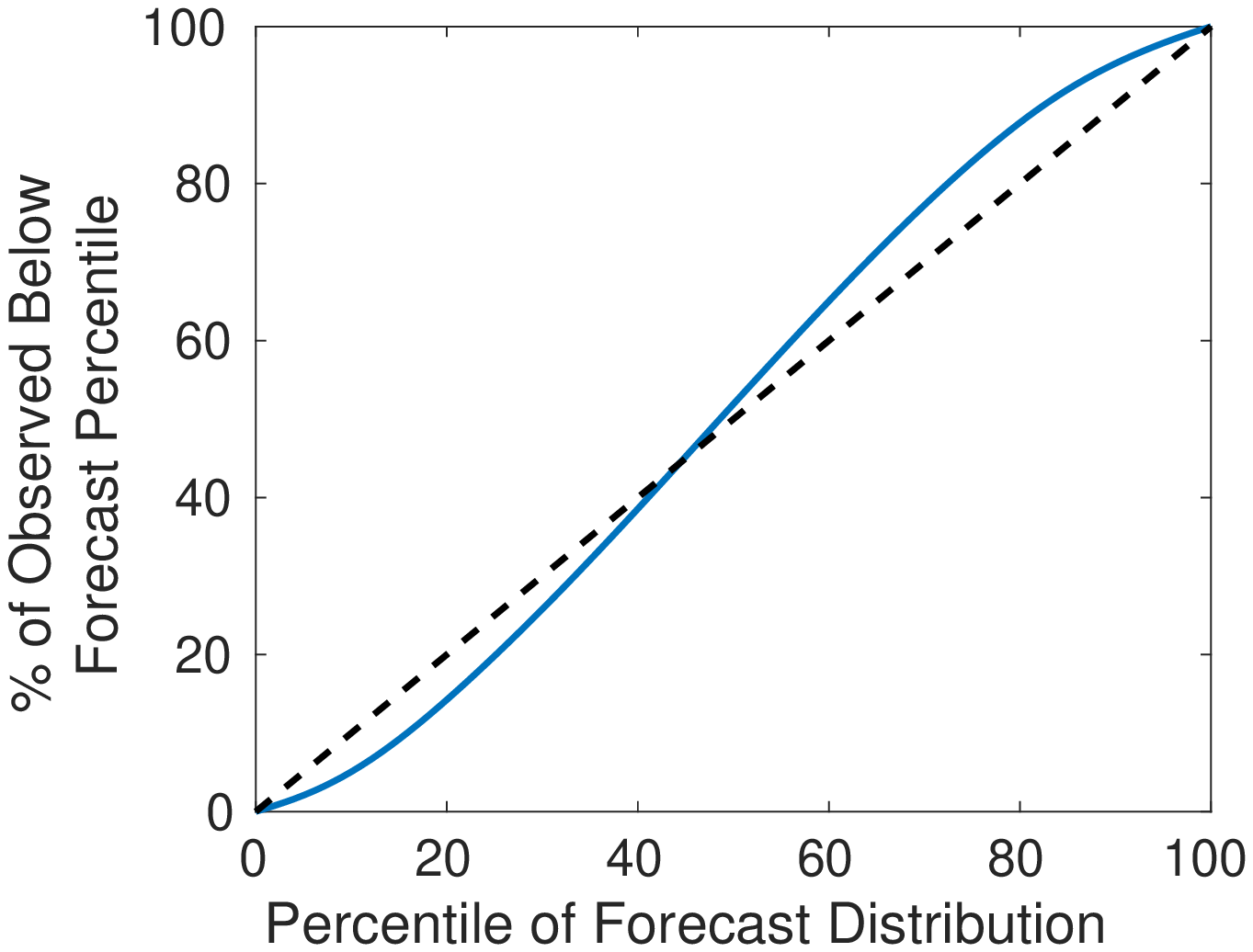}
\caption{Multi-category reliability diagram, as defined in the text. The black dashed line indicates perfect reliability.}
\label{fig:reliability}
\end{figure}

Reliability is an important characteristic of a probabilistic prediction. A prediction is said to be perfectly reliable if for all occasions when a category is predicted with a given probability $p$, the observed frequency of that category is exactly $p$ \citep{gneiting14}. Here, we adopt the definition of reliability for multi-category probabilistic predictions introduced in \citet{hamill97}, which is defined as the average percentage of observations below given percentiles of the predictions distribution. This is calculated as follows. For a single event we calculate the cumulative probability associated to each category. In our case, the probabilistic prediction for a single event is composed of four probabilities.
For instance, for a prediction $(0.3,0.5,0,0.2)$, the cumulative distribution associated to categories $(1,2,3,4)$ is  $(0.3,0.8,0.8,1.0)$. Then, we construct an ordered vector of 100 elements of category numbers $(1,2,3,4)$ at the discrete percentile values $(1,2,\ldots,100)$. This vector represents the predicted category at each percentile. For the above example, this vector would contain 1 for elements 1 to 30, 2 for elements 31 to 80 and 4 for elements 81 to 100. Finally, the reliability, or calibration, for a given percentile is defined as the probability that the observed category is less than the predicted category at this percentile, averaged over all the events. This quantity is shown in Figure \ref{fig:reliability}, where the black dashed line indicates the case of perfect reliability. One can appreciate that the classifier presented in this paper achieves a very good reliability, with predictions slightly under-reliable for percentiles smaller than $50\%$ and slightly over-reliable for percentiles larger than $50\%$. We have verified that the curves using different size of training sets are essentially indistinguishable. 

\subsection{Transition probabilities} 
Once we have assessed that the accuracy and reliability of our classifier are extremely good, we can attempt to classify events that are not labeled, i.e. for which we do not know the category in advance.
Hence, we have classified the whole OMNI dataset (1965-2017), using the entire original dataset described in Section \ref{sec:data} for training. Of course, we can classify only events for which the seven attributes discussed in Section \ref{attributes} are available. Moreover, by considering only events where the largest assigned probability among the four categories is larger than $30\%$, we end up with 292,798 classified events.
At this point, we can compute transition probabilities, defined as the {empirical} probability that a given category at a given time will transition to a different category at later times (or remain in the same category). We have used the hourly OMNI dataset, which however contains gaps that can last for several hours. Therefore, we have only considered as valid the transitions that occur within 10 hours. This slightly reduces the dataset to 290,602 events. Table \ref{table:transition} shows the transition probabilities from the category listed in the first column to the ones listed in the first row. The last column indicates the total number of events for transitions from a given category.

\begin{table}
\caption{Transition probabilities (in percentage)}\label{table:transition}
\centering
\begin{tabular}{c c c c c c}
\hline
from / to & Ejecta & coronal hole & sector reversal & streamer belt & Total number\\
\hline
ejecta & 83.2 &     2.6 &    4.7 &    9.5 & 40,125\\
coronal hole &    1.5 &   94.3 &    0.6 &    3.6 & 58,388\\
sector reversal &    2.7 &    0.5 &   91.7 &    5.1 & 80,515\\
streamer belt &    3.3 &    1.8 &    4.0 &   91.0 & 111,574\\
\end{tabular}
\end{table}

The persistent character of the solar wind is evident: the probability to remain in the same category is always much larger than the probability to transition. However, Table \ref{table:transition} provides more information: when a transition does occur, some transitions are preferred. For instance, it is clear that ejecta transition to streamer belt origin as twice as frequently as they do towards sector reversal origin plasma, and almost four times more frequently than towards coronal hole origin. The transition probabilities of Table \ref{table:transition} have a straightforward physical interpretation. Coronal holes on the Sun are surrounded by streamer belts, hence it makes sense that their transition probability to streamer belts is higher than towards other categories. Also, helmet streamers are surrounded by streamer belts, which explain the higher transition probability between sector reversal region and streamer belt plasma. 
An interesting feature of Table \ref{table:transition} is its asymmetry. {Because the four categories occur with different frequencies, a given transition (excluding self-transitions) might be preferred simply due to such different frequencies}. However, it is straightforward to verify that the asymmetry of Table \ref{table:transition} cannot entirely be due to this effect. Hence, this is a feature that deserves further investigation.
One might wonder if differences of a few percentage points in Table \ref{table:transition} are statistically significant. Hoeffding's inequality \citep{hoeffding63} states that the probability that the empirical mean calculated from a finite independent sample differs from the real expected value is larger than a value $\xi$ can be bounded by $2\exp(-2N\xi^2)$, where $N$ is the {number of independent samples}. In our case, we can estimate that the probability that the values listed differ from the 'real' values by more than $1\%$ is smaller than $0.1\%$ for Ejecta (which is the category with the smaller number of events), and even smaller than $0.01\%$ for the other categories.
Because these transition probabilities are calculated over many years and several solar periods they should not be intended as predictive. However, we believe they represent a very useful baseline that new forecast models should compare against when assessing their accuracy.

\section{Example: one solar rotation in 2006}
{In Figure 3 the Gaussian Process categorization scheme is compared with the \citet{xu2015} algebraic categorization scheme. 27 days of data from the year 2006 are plotted, which is approximately one solar rotation. (These 27 days can also be seen in Fig. 7 of \citet{xu2015}.) The solar wind velocity from the 1-hr resolution OMNI data set is plotted as the black points. This is a time interval when the Earth's footpoint on the rotating Sun passes from a coronal-hole region with an away magnetic orientation (Days 153 and 154) across a helmet streamer into a second coronal-hole region with a toward magnetic orientation (Days 158 - 161) and across another helmet streamer into a third coronal-hole region with an away magnetic orientation (Days 166 - 169). In the bottom portion of Figure  the solar-wind categorizations of the 1-hr-resolution OMNI data set are plotted: the bottom row is the \citet{xu2015} categorization of OMNI, the middle row is the Gaussian Process categorization of OMNI with the maximum vector component required to be greater than 0.3, and the top row is the Gaussian Process categorization of OMNI with the maximum vector component required to be greater than 0.5. The color codings are red for a coronal-hole-origin categorization, green for a streamer-belt-origin categorization, purple for a sector-reversal-region categorization, and blue for and ejecta categorization. As can be seen by comparing the three rows in Figure 3, the Gaussian Process and the Xu and Borovsky schemes agree quite well in the categorizations. The magnetic sector reversals are marked with orange vertical dashed lines; these sector reversals were not determined from the OMNI data set but were instead determined by the direction of the solar-wind electron heat flux measured by the SWEAPM instrument \citep{mccomas98} on the ACE spacecraft in the solar wind. An important test of the categorization is that the sector reversals lie in sector-reversal-region plasma, which they do in Figure 3. Note that the top row is sparser than the bottom two rows: the Gaussian Process $> 0.5$ scheme only yields categorizations that are quite certain in the probabilistic sense. In the year 2006 the Xu+Borovsky scheme categorized 99.4\% of all hours in the year (only not categorizing when there is no solar-wind data available), the Gaussian Process $> 0.3$ scheme categorized 98.4\% of the hours in 2006, and the Gaussian Process $> 0.5$ scheme categorized 42.7\% of the hours in 2006. Note in Figure 3 that the Gaussian Process $> 0.5$ scheme in the top row produces very few stream-belt-origin plasmas categorizations (green). In the year 2006, the Xu and Borovsky scheme had 37.0\% of all categorizations being streamer-belt-origin plasma, the Gaussian Process $> 0.3$ scheme had 25.3\% of all categorizations being streamer-belt-origin plasma, but the Gaussian Process $> 0.5$ scheme had only 11.4\% of all categorizations being streamer-belt-origin plasma. Looking at Figs. 3, 4, and 6 of \citet{xu2015} it is seen that the streamer-belt-origin plasma lies parametrically between the coronal-hole-origin plasma and the sector-reversal-region plasma, and that it is somewhat rare to have plasma that is certainly of that middle category.}

\begin{figure}[ht!]
 \centering
 \includegraphics[width=30pc]{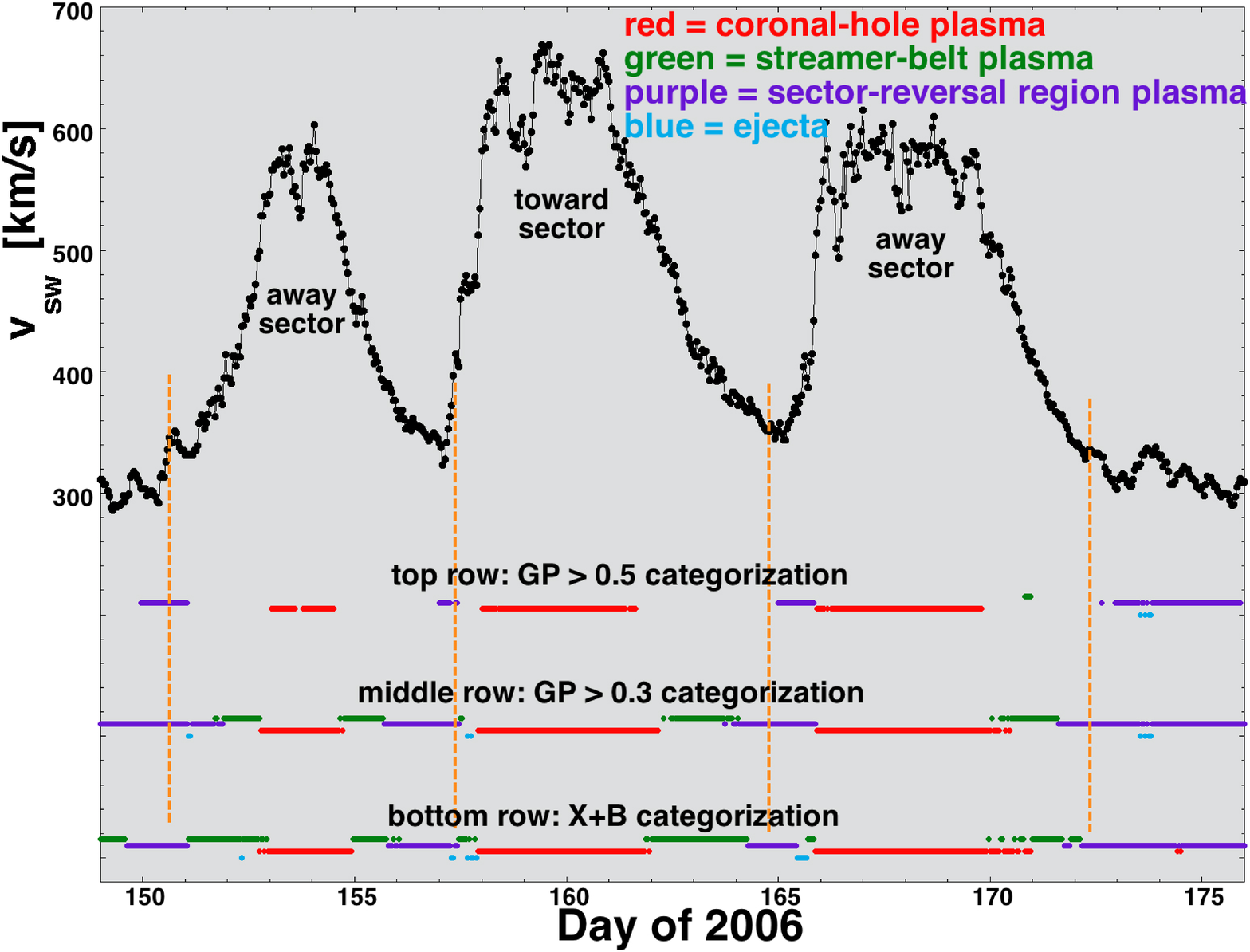}
\caption{For 27 days in 2006 the solar-wind velocity is plotted (black). The times at which magnetic sector reversals in the solar wind pass the Earth are marked by vertical orange dashed lines. In color, categorizations of the OMNI data set are plotted: the bottom row is the \citet{xu2015} categorization, the middle row is the Gaussian Process categorization for the maximum vector component exceeding 0.3, and the top row is the Gaussian Process categorization for the maximum vector component exceeding 0.5.}
\label{fig:figure_3}
\end{figure}

\section{Conclusions}
We have presented a seven-input solar wind classifier based on the technique of Gaussian Processes that yields extremely good results for the four-plasma categorization scheme introduced by \citet{xu2015}. Even using as little as $10\%$ of the dataset for training, the median accuracy is larger than $90\%$ for all categories. This increases to $>96\%$ when the training set is $25\%$ of the original dataset, achieving a peak of $99.7\%$ for the category coronal hole. 
The GP method produces a probabilistic classification. Hence, one can in principle select nowcasts above a given threshold, and classify as 'undecided' the ones below the threshold. We have investigate the ROC curve for varying thresholds, confirming that both the True and False Positive Ratios are optimal for thresholds about 30-40\% 
Finally, we have classified a large portion of the hourly OMNI dataset and studied the probability for a given category to transition to a different category. We have found that, although persistence (i.e. not transitioning) is largely dominant across all categories, transition has a preferential directions: some transition probabilities are several times larger than others.


%

\acknowledgments
The OMNI data is available at https://omniweb.gsfc.nasa.gov/.
The results of the classification applied to the OMNI database (200k+ hours from 1965 to 2017) and the MATLAB code that performs the algorithm presented in this paper are available on the website www.mlspaceweather.org 
\bibliographystyle{agufull08}

\end{article}

%
%
%
%
%

\begin{thebibliography}{27}
\providecommand{\natexlab}[1]{#1}
\expandafter\ifx\csname urlstyle\endcsname\relax
  \providecommand{\doi}[1]{doi:\discretionary{}{}{}#1}\else
  \providecommand{\doi}{doi:\discretionary{}{}{}\begingroup
  \urlstyle{rm}\Url}\fi

\bibitem[{\textit{Bishop}(2007)}]{bishop2007}
Bishop, C. (2007), Pattern recognition and machine learning (information
  science and statistics), 1st edn. 2006. corr. 2nd printing edn,
  \textit{Springer, New York}.

\bibitem[{\textit{Borovsky}(2012{\natexlab{a}})}]{borovsky12}
Borovsky, J.~E. (2012{\natexlab{a}}), Looking for evidence of mixing in the
  solar wind from 0.31 to 0.98 au, \textit{Journal of Geophysical Research:
  Space Physics}, \textit{117}(A6).

\bibitem[{\textit{Borovsky}(2012{\natexlab{b}})}]{borovsky12b}
Borovsky, J.~E. (2012{\natexlab{b}}), The velocity and magnetic field
  fluctuations of the solar wind at 1 au: Statistical analysis of fourier
  spectra and correlations with plasma properties, \textit{Journal of
  Geophysical Research: Space Physics}, \textit{117}(A5).

\bibitem[{\textit{Borovsky}(2016)}]{borovsky16}
Borovsky, J.~E. (2016), The plasma structure of coronal hole solar wind:
  Origins and evolution, \textit{Journal of Geophysical Research: Space
  Physics}, \textit{121}(6), 5055--5087.

\bibitem[{\textit{Chen et~al.}(1997)\textit{Chen, Cargill, and
  Palmadesso}}]{chen97}
Chen, J., P.~J. Cargill, and P.~J. Palmadesso (1997), Predicting solar wind
  structures and their geoeffectiveness, \textit{Journal of Geophysical
  Research: Space Physics}, \textit{102}(A7), 14,701--14,720.

\bibitem[{\textit{Crooker et~al.}(1977)\textit{Crooker, Feynman, and
  Gosling}}]{crooker77}
Crooker, N., J.~Feynman, and J.~Gosling (1977), On the high correlation between
  long-term averages of solar wind speed and geomagnetic activity,
  \textit{Journal of Geophysical Research}, \textit{82}(13), 1933--1937.

\bibitem[{\textit{Gneiting and Katzfuss}(2014)}]{gneiting14}
Gneiting, T., and M.~Katzfuss (2014), Probabilistic forecasting, \textit{Annual
  Review of Statistics and Its Application}, \textit{1}, 125--151.

\bibitem[{\textit{Gopalswamy et~al.}(2015)\textit{Gopalswamy, Yashiro, Xie,
  Akiyama, and M{\"a}kel{\"a}}}]{gopalswamy15}
Gopalswamy, N., S.~Yashiro, H.~Xie, S.~Akiyama, and P.~M{\"a}kel{\"a} (2015),
  Properties and geoeffectiveness of magnetic clouds during solar cycles 23 and
  24, \textit{Journal of Geophysical Research: Space Physics},
  \textit{120}(11), 9221--9245.

\bibitem[{\textit{Hamill}(1997)}]{hamill97}
Hamill, T.~M. (1997), Reliability diagrams for multicategory probabilistic
  forecasts, \textit{Weather and forecasting}, \textit{12}(4), 736--741.

\bibitem[{\textit{Hoeffding}(1963)}]{hoeffding63}
Hoeffding, W. (1963), Probability inequalities for sums of bounded random
  variables, \textit{Journal of the American statistical association},
  \textit{58}(301), 13--30.

\bibitem[{\textit{Kilpua et~al.}(2016)\textit{Kilpua, Madjarska, Karna,
  Wiegelmann, Farrugia, Yu, and Andreeova}}]{kilpua16}
Kilpua, E., M.~Madjarska, N.~Karna, T.~Wiegelmann, C.~Farrugia, W.~Yu, and
  K.~Andreeova (2016), Sources of the slow solar wind during the solar cycle
  23/24 minimum, \textit{Solar Physics}, \textit{291}(8), 2441--2456.

\bibitem[{\textit{MacKay}(1996)}]{mackay96}
MacKay, D.~J. (1996), Bayesian methods for backpropagation networks, in
  \textit{Models of neural networks III}, pp. 211--254, Springer.

\bibitem[{\textit{McComas et~al.}(1998)\textit{McComas, Bame, Barker, Feldman,
  Phillips, Riley, and Griffee}}]{mccomas98}
McComas, D., S.~Bame, P.~Barker, W.~Feldman, J.~Phillips, P.~Riley, and
  J.~Griffee (1998), Solar wind electron proton alpha monitor (swepam) for the
  advanced composition explorer, in \textit{The Advanced Composition Explorer
  Mission}, pp. 563--612, Springer.

\bibitem[{\textit{Neal}(2012)}]{neal12}
Neal, R.~M. (2012), \textit{Bayesian learning for neural networks}, vol. 118,
  Springer Science \& Business Media.

\bibitem[{\textit{Neugebauer et~al.}(2016)\textit{Neugebauer, Reisenfeld, and
  Richardson}}]{neugebauer16}
Neugebauer, M., D.~Reisenfeld, and I.~G. Richardson (2016), Comparison of
  algorithms for determination of solar wind regimes, \textit{Journal of
  Geophysical Research: Space Physics}, \textit{121}(9), 8215--8227.

\bibitem[{\textit{Rasmussen and Nickisch}(2010)}]{rasmussen10}
Rasmussen, C.~E., and H.~Nickisch (2010), Gaussian processes for machine
  learning (gpml) toolbox, \textit{Journal of Machine Learning Research},
  \textit{11}(Nov), 3011--3015.

\bibitem[{\textit{Rasmussen and Williams}(2006)}]{rasmussen2006}
Rasmussen, C.~E., and C.~K. Williams (2006), Gaussian processes for machine
  learning. 2006, \textit{The MIT Press, Cambridge, MA, USA}, \textit{38},
  715--719.

\bibitem[{\textit{Riley et~al.}(2017)\textit{Riley, Ben-Nun, Linker, Owens, and
  Horbury}}]{riley17}
Riley, P., M.~Ben-Nun, J.~A. Linker, M.~Owens, and T.~Horbury (2017),
  Forecasting the properties of the solar wind using simple pattern
  recognition, \textit{Space Weather}, \textit{15}(3), 526--540.

\bibitem[{\textit{Salem et~al.}(2003)\textit{Salem, Hubert, Lacombe, Bale,
  Mangeney, Larson, and Lin}}]{salem03}
Salem, C., D.~Hubert, C.~Lacombe, S.~Bale, A.~Mangeney, D.~Larson, and R.~Lin
  (2003), Electron properties and coulomb collisions in the solar wind at 1 au:
  Wind observations, \textit{The Astrophysical Journal}, \textit{585}(2), 1147.

\bibitem[{\textit{Stakhiv et~al.}(2015)\textit{Stakhiv, Landi, Lepri, Oran, and
  Zurbuchen}}]{stakhiv15}
Stakhiv, M., E.~Landi, S.~T. Lepri, R.~Oran, and T.~H. Zurbuchen (2015), On the
  origin of mid-latitude fast wind: Challenging the two-state solar wind
  paradigm, \textit{The Astrophysical Journal}, \textit{801}(2), 100.

\bibitem[{\textit{Williams and Barber}(1998)}]{williams98}
Williams, C.~K., and D.~Barber (1998), Bayesian classification with gaussian
  processes, \textit{IEEE Transactions on Pattern Analysis and Machine
  Intelligence}, \textit{20}(12), 1342--1351.

\bibitem[{\textit{Wimmer-Schweingruber
  et~al.}(2006)\textit{Wimmer-Schweingruber, Crooker, Balogh, Bothmer, Forsyth,
  Gazis, Gosling, Horbury, Kilchenmann, Richardson et~al.}}]{wimmer06}
Wimmer-Schweingruber, R., N.~Crooker, A.~Balogh, V.~Bothmer, R.~Forsyth,
  P.~Gazis, J.~Gosling, T.~Horbury, A.~Kilchenmann, I.~Richardson, et~al.
  (2006), Understanding interplanetary coronal mass ejection signatures,
  \textit{Space Science Reviews}, \textit{123}(1-3), 177--216.

\bibitem[{\textit{Wing et~al.}(2016)\textit{Wing, Johnson, Camporeale, and
  Reeves}}]{wing16}
Wing, S., J.~R. Johnson, E.~Camporeale, and G.~D. Reeves (2016), Information
  theoretical approach to discovering solar wind drivers of the outer radiation
  belt, \textit{Journal of Geophysical Research: Space Physics},
  \textit{121}(10), 9378--9399.

\bibitem[{\textit{Winterhalter et~al.}(1994)\textit{Winterhalter, Smith,
  Burton, Murphy, and McComas}}]{winterhalter94}
Winterhalter, D., E.~Smith, M.~Burton, N.~Murphy, and D.~McComas (1994), The
  heliospheric plasma sheet, \textit{Journal of Geophysical Research: Space
  Physics}, \textit{99}(A4), 6667--6680.

\bibitem[{\textit{Xu and Borovsky}(2015)}]{xu2015}
Xu, F., and J.~E. Borovsky (2015), A new four-plasma categorization scheme for
  the solar wind, \textit{Journal of Geophysical Research: Space Physics},
  \textit{120}(1), 70--100.

\bibitem[{\textit{Zurbuchen et~al.}(2002)\textit{Zurbuchen, Fisk, Gloeckler,
  and Von~Steiger}}]{zurbuchen02}
Zurbuchen, T., L.~Fisk, G.~Gloeckler, and R.~Von~Steiger (2002), The solar wind
  composition throughout the solar cycle: A continuum of dynamic states,
  \textit{Geophysical research letters}, \textit{29}(9).

\bibitem[{\textit{Zurbuchen}(2007)}]{zurbuchen07}
Zurbuchen, T.~H. (2007), A new view of the coupling of the sun and the
  heliosphere, \textit{Annu. Rev. Astron. Astrophys.}, \textit{45}, 297--338.

\end{thebibliography}
%
%
%
%





\end{document}